\documentclass[
aps,pra,
reprint,
superscriptaddress,
amsmath,
amssymb
]{revtex4-1}

\usepackage[utf8]{inputenc}
\usepackage{times}
\usepackage{microtype}
\usepackage{graphicx}
\usepackage{upgreek}
\usepackage{bm}
\usepackage{hyperref}
\usepackage{xcolor}
\usepackage[normalem]{ulem}
\usepackage{placeins}
\usepackage{dcolumn}

\definecolor{jens}{rgb}{0,0,0}
\definecolor{mjk}{rgb}{0,0,0}
 \definecolor{mh}{rgb}{0,0,0}
 \definecolor{etc}{rgb}{0,0,0}

 \definecolor{jens}{rgb}{0,0,0}
\newcommand{\jen}[1]{{\color{jens}#1}}
\newcommand{\mjk}[1]{{\color{mjk}#1}}
\newcommand{\etc}[1]{{\color{etc}#1}}
\newcommand*{\mh}[1]{{\color{mh}#1}}

\hypersetup{colorlinks=true,allcolors=black,urlcolor=blue}

\hypersetup{
	pdfauthor={Michael Herold, Michael J. Kastoryano, Earl T. Campbell, Jens Eisert},
	pdftitle={Fault tolerant dynamical decoders for topological quantum memories}
}

\renewcommand*{\vec}[1]{\ensuremath{\mathbf{#1}}}

\DeclareMathOperator{\polylog}{polylog}

\newcommand{\VV}{\mathbb{V}}

\newcommand{\CCC}{\mathcal{C}}

\newcommand{\EEE}{\mathcal{E}}

\newcommand{\LLL}{\mathcal{L}}

\begin{document}
\title{Cellular automaton decoders of topological quantum memories in the fault tolerant setting}

\author{Michael\ Herold}

\affiliation{Dahlem Center for Complex Quantum Systems, Freie Universit{\"a}t Berlin, 14195 Berlin, Germany}

\author{Michael J.\ Kastoryano}
\affiliation{NBIA, Niels Bohr Institute, University of Copenhagen, 2100 Copenhagen, DK.}

\author{Earl\ T.\ Campbell} 

\affiliation{Department of Physics and Astronomy, University of Sheffield, Sheffield S3 7RH, UK}

\author{Jens\ Eisert}
\affiliation{Dahlem Center for Complex Quantum Systems, Freie Universit{\"a}t Berlin, 14195 Berlin, Germany}

\begin{abstract}
Active error decoding and correction of topological quantum codes -- in particular the toric code -- remains one of the most viable routes to large scale quantum information processing. In contrast, passive error correction relies on the natural physical dynamics of a system to protect encoded quantum information. However, the search is ongoing for a completely satisfactory passive scheme applicable to locally-interacting two-dimensional systems. Here, we investigate dynamical decoders that provide passive error correction by embedding the decoding process into local dynamics. We propose a specific discrete time cellular-automaton decoder in the fault tolerant setting and provide numerical evidence showing that the logical qubit has a survival time extended by several orders of magnitude over that of a bare unencoded qubit. We stress that (asynchronous) dynamical decoding gives rise to a Markovian dissipative process. We hence equate cellular-automaton decoding to a fully dissipative topological quantum memory, which removes errors continuously. In this sense, uncontrolled and unwanted local noise can be corrected for by a controlled local dissipative process. We analyze the required resources, commenting on additional polylogarithmic factors beyond those incurred by an ideal constant resource dynamical decoder.
 \end{abstract}

\maketitle

\subsection{Introduction}
Quantum coherence is capricious, and taming it requires sophisticated methods of control and significant resource overhead. So far the best candidates are based on Kitaev's toric code \cite{kitaev97,DKL02}, and variants thereof. While quantum information can be stored in ground states of a toric Hamiltonian, thermal excitations are unconfined in two or three dimensions, allowing thermal diffusion and corrupting data in constant time~\cite{Nussinov08,HCC09,AFH09,LYP15,bardyn15}. Using active decoding, this problem can be overcome.  Errors must be regularly monitored by collecting syndrome information, which is analyzed by a decoding algorithm and corrected for. The first steps towards experimental implementations of error-correcting protocols are under way \cite{Hanson14,Wrachtrup14,Sun14,Martinis15}.  

Given reliable snapshots of error syndromes, decoding algorithms can be independently applied to each time slice \cite{TerhalReview}.  Decoders exist using notions of minimum-weight perfect matching~\cite{WHP03,FWH12b}, renormalisation~\cite{DP10b,BH13}, and simulated annealing~\mh{\cite{WL12,HWL14}}, which is highly  problematic for 2D architectures, where on chip wiring becomes cumbersome beyond 9 physical qubits.  However, parallelized decoding is also possible with cellular-automaton decoders that locally store syndrome information, communicating with only nearest neighbors.  Each node in the cellular automaton stores a limited amount of additional data, and our focus here is on so-called $\phi$-automata~\cite{HCEK14} where this data is a single variable, or $\phi$-value, per node. 

Yet, perfect syndrome measurement is a fantasy.  It suffices for developing toy decoders, but is merely the first step to practical decoding.  Conventionally, one tackles measurement errors by recording a history of recent syndrome snapshots, then running a decoder suitable for three-dimensional syndrome distributions~\cite{WHP03,FWH12b}. 

In contrast, here we propose the concept of a Markovian dynamical decoder, where each step of the decoder only depends on the present configuration of (faulty) syndromes measurements and the auxiliary $\phi$-field.  Specifically, we show that a $\phi$-automaton decoder can tolerate measurement noise without significant changes to the algorithms, while preserving a number of desirable features.  Measurement errors cause ripples in the dynamics, but may be naturally tolerated, provided fresh syndrome information is acquired before the ripple travels too far. No explicit record of the syndrome history is required, rather past syndromes merely leave echoes in the state of the cellular automaton.  To the best of our knowledge, there is only one prior study of a Markovian dynamical decoder; due to Harrington~\cite{Har04}.   Interest in this early work has recently been revived~\footnote{\emph{Private communication}, B.\ Terhal (2016).}, and appears to be the right setting for certain types of non-Abelian anyon decoding~\cite{HutterHDRG,PoulinAnyons}. 

\begin{figure}\centering
\includegraphics{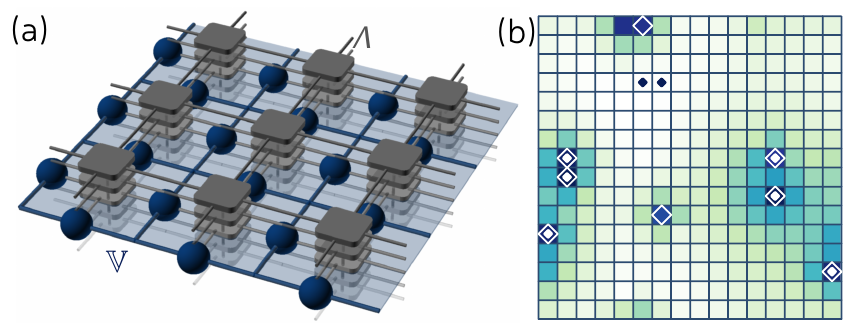}
\caption{\textbf{(a)} 
Layout of the decoder. Spheres represent toric code qubits and gray cells are interconnected elements of the cellular automaton, extending into the third dimension. \textbf{(b)} Snapshot of the asynchronous decoder on the toric code plane for a given numerical run. Filled diamonds represent actual anyons, empty diamonds correspond faulty measurement that incorrectly predict the presence of an anyon. Both types of syndromes act as field sources. The color scale represents the field intensity which builds up around anyons.}
\label{fig:schema}
\end{figure}

The Markovian dynamical decoder paradigm is naturally suited to various physical implementations where syndromes can not easily be measured synchronously.  In systems where entanglement is generated probabilistically~\cite{CCGZ99,LBK05,BK05,CB08}, including syndrome measurements~\cite{NYB13}, this situation readily applies. Rather than a well defined series of snapshots, the syndromes are refreshed \emph{asynchronously}; meaning that every unit cell of the cellular-automaton decoder is updated with a certain probability that is independent of the other cells.  The dynamical paradigm is equally plausible in other architectures with an inherently modular structure of interacting nodes, as is notably the case in  anticipated mesoscopic implementations of the toric code \cite{TerhalDiVincenzo,Altland}, based on Majorana bound states supported by topological semi-conductor nanowires. 

The locality, Markovianity and asynchronicity of such decoders makes them conceptually equivalent to certain physical systems undergoing open system dynamics. We show that an asynchronous cellular-automaton decoder can be considered as the discretization of a Markovian master equation which dissipatively protects the topological subspace of the code.  This opens up the potential of engineering quantum memories from locally interacting matter systems, providing passive protection against noise. In this sense, one can think of passive dissipative quantum memories protected against unwanted dissipation. Finally, from rudimentary numerical exploration, dynamical cellular-automaton decoders seem promising in the presence of qubit loss or in the presence of spatially and temporally correlated noise, though these settings are all beyond the scope of the present article.

An asynchronous cellular automaton is most naturally built from a synchronous one that does not require a clock or time dependent rules.  One such time homogeneous cellular automaton is provided by the 3D $\phi$-automaton of Ref.~\cite{HCEK14}.  In contrast, the 2D${^*}$ $\phi$-automaton of Ref.~\cite{HCEK14} and an automaton studied by Harrington~\cite{Har04,TerhalInPreparation} both keep track of time, and so are less easily adapted to an asynchronous setting. For completeness, we later review the synchronous operation of the Harrington decoder. We show that our dynamical 3D $\phi$-automaton decoder provides such an asynchronous dynamical decoder.

\section{The decoders}
\subsection{$\phi$-automaton decoders}
Decoders in our context are processes that remove errors from a lattice configuration $\VV$ and bring the system's state back to the code space. Here we focus on the toric code that uses an $L$ by $L$ square lattice with periodic boundary conditions. For simplicity, we assume uncorrelated $X$ and $Z$ Pauli noise. This allows us to analyse only Pauli $X$ errors, which are measured by \emph{plaquette operators}, with $Z$ noise decoding behaving identically and independently. To restrict ourselves to algorithms that can be highly parallelized and are inherently local, we formulate all algorithms as \emph{cellular automata (CA)}. For every measurable syndrome in the code we also place a cell for the automaton, constituting a parallel lattice $\Lambda$ with one \mh{s}ite per plaquette. The cell has local access to the measured syndrome and can read information from the neighboring cells on the CA lattice $\Lambda$ of side length $L$. If a $-1$ syndrome is measured, we will say that an \emph{anyon} is present. The anyons can be moved locally by applying (or equivalently storing) local bit flip operations. The automaton possesses the same periodic boundary conditions as the code does. As we will see, it is sometimes convenient to allow the CA lattice $\Lambda$ to extend in the third dimension (see Fig.~\ref{fig:schema}). 

We specifically focus on and generalize a scheme for CA decoding the 2D toric code, introduced in Ref.~\cite{HCEK14}. The underlying idea is to simulate an attractive interaction between anyons from local update rules.   The attraction is mediated via a scalar field $\phi(\vec x)$ on the discrete lattice $\vec x \in \Lambda$. Sites occupied by anyons act as sources for the field.  A field value is stored at each cell of the automaton, and the field builds up according to a local discretized version of Poisson's equation. As it turns out, the field has to be generated via a CA which extends into the third dimension so as to guarantee that the field falls off as $1/r$ away from  anyons~\cite{HCEK14}. That way, nearby anyons are attracted to each other under the influence of the field, but do not feel the attraction of far away anyons very strongly. Generating such a field locally in two dimensions seems to be difficult, but it occurs naturally in 3D. We will briefly review the so-called 3D $\phi$-automaton depicted in Fig.~\ref{fig:schema}a.  Other passive memories have been proposed where an effective interaction emerges in the system Hamiltonian~\cite{PCL11,PHW13}, but here the interaction is simulated by the dynamics of a cellular automata rather than being a manifestly physical interaction.

The cellular automaton update rules dynamically evolve a combined quantum-classical system. 
For ease of presentation, the field and anyon positions are embedded in quantum systems with Hilbert spaces ${\cal L}$ and ${\cal V}$, respectively. We use $\sigma$ to denote states within the composite space ${\cal H \otimes \cal L\otimes \cal V}$, where ${\cal H}$ is the toric code Hilbert space.
The update rules are composed of two elementary components: field updates $F$ and anyon updates $A$. 
The field update $F$ simultaneously replaces the field value at all sites by the average of all six neighboring field values and adds $+1$ when an anyon is present. The anyon update $A$ moves anyons to the one of their four neighboring sites in the toric code plane with the highest associated $\phi$-value. Unwanted limit cycles are avoided by moving with a probability of $1/2$.  

In order for two anyons to be attracted to each other, they must ``feel'' the field from the other anyon. In other words, the influence of an anyon on the field $\phi$ needs to have propagated to other anyons. To ensure a sufficient convergence of the field between anyon updates, it is required that $F$ is applied $c\sim\log^2 L$ times~\cite{HCEK14}. This yields the compound update rule $D_c := A F^c$. Applying the update $D_c$ several times guarantees that typical clusters of errors get collapsed back to an anyon free configuration.

\subsection{Markovian dynamical decoders}
We next  introduce the notion of a \emph{Markovian dynamical decoder}, for the moment still in the synchronous setting. \mjk{A dynamical decoder is to be contrasted with a regular \jen{
decoder} in that a regular decoder minimizes the number of residual errors at each time slice. 
\jen{At a given instance in time,} a dynamical decoder only chooses to remove some of the error, according to a 
\jen{suitable} prescription. A Markovian dynamical decoder keeps no memory of previous measurement rounds. Dynamical decoders do not need to be local, as in Ref.\ \cite{FWH12b}, but we will only consider local ones in this \jen{work. In fact, dynamical decoding is expected
to be particularly advantageous if constraints of locality are to be respected.}}
 
\jen{We will now turn to making these notions more precise. We} denote by $E_p$ the error map that applies $X$-flips on each qubit of the toric code with probability $p$. The imperfect error syndrome extraction is reflected by the application of a map $M_q$, where $q>0$ denotes the probability that a syndrome is incorrect. For perfect syndrome measurement, $q=0$, decoding can be achieved with only one round of syndrome extraction and a suitable number $\tau$ of updates $D_c$,
\begin{equation}
\sigma_\tau = ( D_c^\tau  M_0  E_p )( \sigma ),   
\label{eq:static}
\end{equation}
where $\sigma$ is a state on the combined quantum-classical Hilbert space.  The quantum component is the toric code and the classical component represents both the $\phi$-field and the anyon locations reported by measurements.  The exponent $\tau$ must be chosen such that the decoder terminates. If the error rate $p$ is below a certain critical threshold, a logical error is returned with probability exponentially suppressed in $L$. However, for active error correction only \emph{partial} decoding is necessary between different rounds of syndrome measurements. In particular, one may consider very short \emph{sequences} in which only partial corrections are calculated before the syndrome is remeasured.  As every recovery step only depends on the instantaneous time syndrome information, we refer to this concept as \emph{Markovian dynamical decoding}. Specifically, we define synchronous dynamical decoding as the succession of the maps
\begin{equation}
 \sigma_\tau = \left(D_c M_q E_p\right)^\tau (\sigma), \label{eq:dynanmic-synchronous}
\end{equation}
where $\tau$ is the number of sequences. In the setting of dynamical decoding, we are interested  in estimating the survival time of the logical subspace; i.e. how long one typically has to wait before a logical $X$ or $Z$ error occurs on the encoded qubits. In order to determine this survival time we check whether a logical error has
occurred by decoding a copy of the state without measurement errors in every sequence. 

It is important to note that between every round of measurement, the dynamical decoder will leave many residual errors in the system. This is even the case when the measurement error goes to zero. This is in stark contrast with conventional decoders designed for perfect measurements, which try to minimize the residual errors in the lattice between every measurement. A code is {called} \emph{single-shot error correcting}, if the residual errors can be minimized at each step in the presence of measurement errors, while still guaranteeing a decoding 
threshold {\cite{Bom14,BNB15}}. 
Dynamical decoders are not in general perform \etc{single-shot error correction} but the converse is true.

\subsection{Decoding in continuous time}
Equation~\eqref{eq:dynanmic-synchronous} defines synchronous Markovian dynamical decoding, where for the $\phi$-automaton decoder the field updates $F$ and the anyon updates $A$ are applied simultaneously on all sites at each time step. A more physically relevant scheme would be one that is naturally \emph{asynchronous}. We consider the local operators $f_\vec{x}$ and $a_\vec{x}$ performing field and anyon updates but only applied to a single site $\vec{x}$, and the $X_{p,\vec{v}}$-operator applying a Pauli operator to the edge $\vec{v}\in\VV$ with probability $p$. The $m_{q,\vec{x}}$-operator reflects a measurement with error probability $q$ at $\vec{x}\in\Lambda$. We pick a site $\vec{x}$ (or edge $\vec{v}$ in the case of $X_{p,\vec{v}}$) at random, and apply the local map to that site (or edge).  \etc{By considering nonzero $q$, we allow for noisy measurements.  Every site keeps a record of its most recent measurement result, and the field updates are calculated on the basis of this most recent result.  Therefore, field updates are based on a record of anyon locations that is both noisy and not always up-to-date.}  Furthermore, the probability to choose $f_\vec{x}$ is weighted with $c\sim\log^2 L$, \etc{so we incur a polylogarthmic increase in the rate of field updates relative to anyonic and measurement processes}. To compare the performance of synchronous and asynchronous dynamical decoders, we define one time step as the number of operations after which all anyons on average had the chance to move once.

\section{Numerical results}
\label{Sec::NumRes}

\subsection{Results from simulations}

\begin{figure*}
\includegraphics{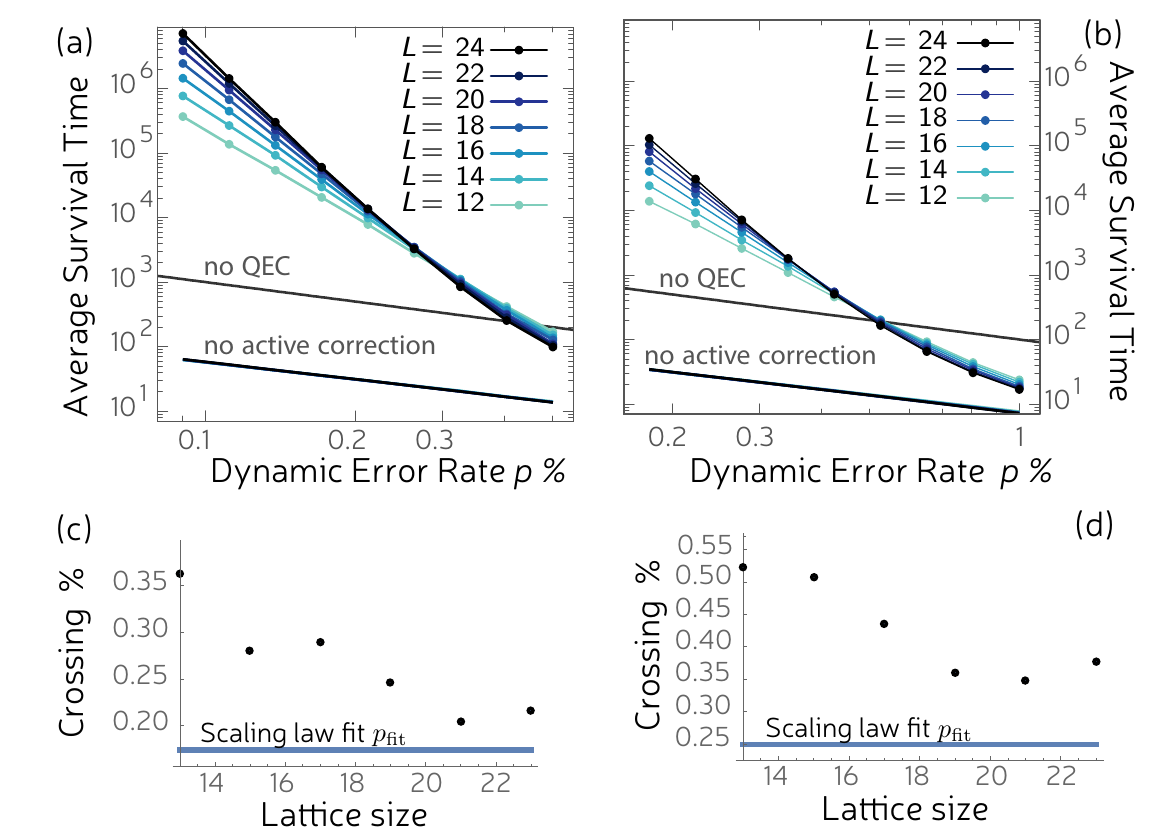}%
\caption{Rapid increase in the average survival time of the dynamical 3D $\phi$-automaton decoder with measurement errors ($q=p$) in  \textbf{(a)} synchronous and \textbf{(b)} asynchronous operation. Below are curve crossings as a function of $p$ in \textbf{(c)} synchronous and \textbf{(d)} asynchronous operation. A data point at lattice size $L$ represents crossing between curves for lattice sizes $L-1$ and $L+1$. Also shown is $p_{\mathrm{fit}}$, the scaling law fit that attempts to compensate for finite size effects.}
\label{fig:results}
\end{figure*}

We have simulated the proposed synchronous and asynchronous Markovian dynamical decoders for even lattice sizes $L=12,14,\ldots, 24$ by Monte Carlo sampling.  We allowed for measurement errors $q=p$ and considered a range of physical error probabilities.  Fig.~\ref{fig:results}a presents results for the synchronous dynamical decoder, and Fig.~\ref{fig:results}b presents the asynchronous case. \mjk{Each data point \jen{reflects} 
the average over 5000 samples. } For both decoders, we see clear evidence of survival times extended by several orders of magnitude across the parameter range studied.  Typically, one aims to numerically identify a threshold, below which survival times are exponentially prolonged with increasing lattice size.  Before discussing possible threshold values, we review the standard techniques for identifying thresholds.

An established methodology is to look for a common crossing between curves for the various lattice sizes.  Yet, in practice, even when a threshold exists, the crossings drift to lower error rates with increasing lattice size.  At sufficiently large lattice sizes, this drift will become negligible and a common crossing is observed and taken as firm evidence of a threshold. A more sophisticated approach is to fit data to a scaling law \cite{WHP03,Har04}
\begin{multline}
\left\langle T(L, p) \right\rangle = 
A +
B \left(p-p_\text{fit}\right) L^{1/\nu} +
\\ +
C \left(p-p_\text{fit}\right)^2 L^{2/\nu} +
D \, L^{-1/\mu} \,
\label{eq:fit}
\end{multline}
where the term $D \, L^{-1/\mu}$ accounts for finite lattice size effects.  The variable 
$p_\text{fit}$ gives a threshold estimate and will be smaller than the lowest observed crossing, having extrapolated to account for finite size effects.  

Our numerical investigations show crossings in the range $0.22 \% -  0.36 \%$ for the synchronous decoder and for the $0.34\% - 0.52\%$ asynchronous case.  The lowest crossings at $p_{\text{cross}}=0.22 \%$ (synchronous) and $p_{\text{cross}}=0.30 \%$ (asynchronous) correspond \etc{the} crossing between the $L=22$ and $L=24$ curves and give threshold upper bounds.  Next, we select the data with $p$ close to these lowest crossings.  Specifically, for the synchronous decoder we select data points with $p=\{ 0.18\% , 0.223\%, 0.2763\% \}$ and for the asynchronous case we select data points with $p=\{ 0.223\%, 0.2763\%, 0.3424 \% \}$.  Fitting this data to the above scaling law, gives threshold estimates of $p_\text{fit}=0.17 \%$ (synchronous) and $p_\text{fit}=0.25 \%$ (asynchronous).  As expected, these $p_\text{fit}$ values lie below the corresponding lowest crossing $p_{\text{cross}}$.  However, they are sufficiently far below the lowest crossing that we can not conclusively claim to have found a threshold at the value $p_\text{fit}$.  

We conclude that finite size effects are still prominent at $L=24$.  The use of the scaling law fitting is intended to adjust for such effects, but can not be relied upon for large adjustments.  Ideally, data at higher $L$ would be available, but we are already at the limits of available computational power.  Other decoders have been able to collect reliable data at higher $L$.  While cellular automata can be expected to run quickly on custom hardware, our simulations were executed on a conventional serial processor, leading to an $O(L^3)$ slow down in the runtime, which constrains the range of accessible $L$ values.  The need to sample at small probabilities and long survival times, also leads to great computational overhead since more samples are required to obtain good statistics.  

It is important to note that $\phi$-field decoders suffer very significant finite size effects up to large system sizes.  Evidence of this is found in Sec.~\ref{Sec::Instant} where we investigate a related class of decoders that behave very similarly, but which can reach higher $L$ values on a conventional computer architecture.  This opens the possibilities of inferring higher $L$ performance for our CA decoders by bootstrapping our numerical data onto data provided by these related decoders.  However, we again found that finite size effects were visibly clear upto $L=50$, and even upto $L=80$ did not provide a scaling law fit $p_\text{fit}$ that was comparable to the lowest observed crossings.

These difficulties seem to be generic for Markovian dynamical decoders so far proposed. For instance, the dynamical decoder identified by Harrington~\cite{Har04} is believed to have a threshold by virtue of an extension of the work of G{\'a}cs \cite{Gac86,Gac01,Gra01}.  However, all numerical investigations~\cite{Har04,TerhalInPreparation} to date have yet to find a common crossing over a wide range of lattice sizes.

\subsection{Measurement errors}
For many of the existing topological quantum codes, it is necessary to keep a log of the syndrome information whenever there are measurement errors. Only codes allowing for \emph{single shot decoding} \cite{Bom14,BNB15} can cope with measurement errors in the setting of Eq.~(\ref{eq:static}). One of the main benefits of both synchronous and asynchronous dynamical decoders is that measurement errors are naturally incorporated  without any additional overhead. This feature is inherited from the local structure of the decoding algorithm: If the presence of an anyon is falsely indicated, this may only result in a single additional bit-flip error created by the local anyon move. In the next sequence this bit flip is likely incorporated into the syndrome and is not different from usual errors. Also, the imaginary anyon may attract other anyons, which is not different from the case where a real error occurred. 

\subsection{Instantly updated $\phi$-decoders}
\label{Sec::Instant}

The simulation has to run throughout the full survival time, which grows very rapidly with increasing system size and decreasing error rates.  Therefore, the simulation of dynamical 3D $\phi$-automaton decoders on traditional architecture is numerically expensive.  Furthermore, we simulate the 3D cellular automata within  a serial centralized computing model, which leads to an additional time cost scaling with $L^3$.  Parallelisation exploiting the cellular nature of the decoder would eliminate this cost, but has not yet been implemented.  Accordingly, the simulation is clearly limited in system size to low error rates. 

\begin{figure}
\includegraphics{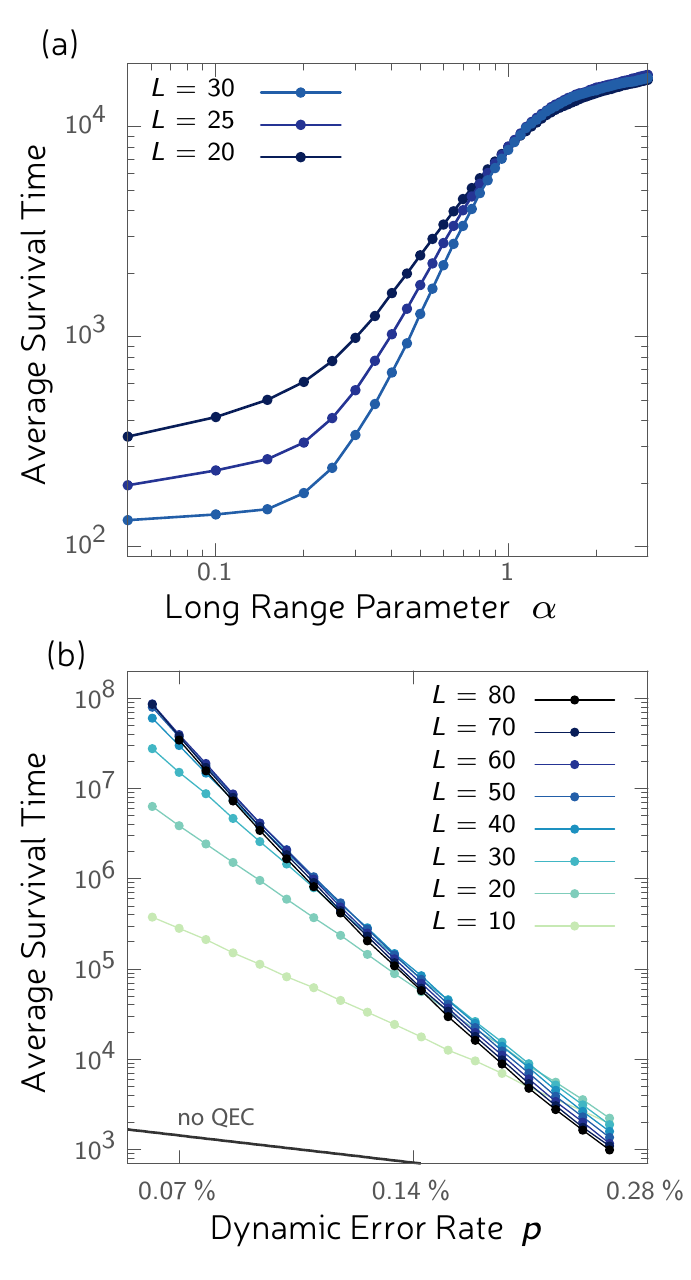} 

\caption{Details of dynamical decoders simulated with explicit fields with  $1/r^\alpha$ profile and without measurement errors. \textbf{(a)} Survival time for different values of $\alpha$ with $p=0.6\,\%$.  \textbf{(d)} The $1/r^{1.05}$-decoder shows survival time increasing as a function of system size, but with crossings drifting to lower error rates.}
\label{fig:appendix-1overR-details}
\end{figure}

To obtain results for larger system sizes we replace the $\phi$-automaton by explicitly calculating steady-state field values. The appropriate field for a 3D $\phi$-automaton is a superposition of fields decaying as $1/r$, where $r$ represents the Manhattan, or 1-norm, distance from the anyon position. The anyon-move rules are identical to the synchronous CA decoder.  The dimensionality of a $\phi$-automaton determines for which number of dimensions the discretized Poisson equation is solved through the iterative algorithm. To construct a $\phi$-automaton decoder with a time independent parameter $c$, it seems necessary to choose a three dimensional $\phi$-automaton~\cite{HCEK14}. Fig.~\ref{fig:appendix-1overR-details}a gives the results for a dynamical decoder that operates with explicitly calculated field values $1/r^{\alpha}$ for different choices of the parameter $\alpha$. We find that the parameter $\alpha$ has a strong impact on the survival time. An increase of survival time with lattice size is only present for $\alpha \geq 1$.  The iterative field generation via a CA leads to incomplete convergence of the field leading to fields with slightly shorter range. To incorporate this fact we opted to study an explicit field decaying as $1/r^{1.05}$. For further details on the field profile and convergence properties we refer to Ref.~\cite{HCEK14}.

As expected, the explicit field $1/r^{1.05}$ decoder and the 3D CA decoder show similar behaviour.  Numerics up to system size $L=80$ can be found in Fig.~\ref{fig:appendix-1overR-details}b. Fitting Eq.~\eqref{eq:fit} we obtain threshold values $p_\text{fit}$ from 0.03\,\% to 0.05\,\% with measurement errors $q=p$. The range is obtained by excluding different amounts of small system size data and varying the used error ranges $p$. However, all our data and observed crossing are for $0.07 \%$ and above.  Since $p_\text{fit}$ is again extrapolated to below the regime where we have concrete data, we are again unable to decisively conclude that $p_\text{fit}$ is an accurate threshold estimate. In other words, Fig.~\ref{fig:appendix-1overR-details}b tells us that if there is a threshold, then it must be below $0.07\%$. Since the crossings for larger lattice sizes happen at average survival times of $\sim 10^8$ time steps, we cannot directly probe the behavior below threshold, and cannot guarantee that there is exponential increase in survival time as a function of system size below a critical $p$.  Fig.~\ref{fig:appendix-1overR-details}b is consistent with other numerical studies of markovian dynamical decoders in Refs.\ \cite{Har04,TerhalInPreparation}.


\section{Further conceptual characteristics of Markovian dynamical decoders}

\subsection{Decoding by open systems dynamics}

The concept of dynamical decoders is motivated by the theory of classical robust memories. In this context, cellular automata (CA) are studied that redundantly store the information in their local state space. Appropriately tailored update rules can correct local errors and can be used for active error correction. Such error correcting features appear in two fundamental varieties. First, there exist automata which are called \emph{eroders} or \emph{washout automata}. Considering a binary CA, the repeated action of an eroder leads to all cells having the value of the initial states majority occupation. In this way, eroders are the equivalent of a non-dynamical decoder in the absence of measurement errors. Both systems return a partially erroneous memory state to an error free state. It is suggestive to imagine that in this setting the errors only act at the first time slice and afterwards the eroder can act unimpeded. The second variety appears if we let errors appear at every single time slice. In this setting the following question is asked: Given the initial state is a fixed point of the automaton, is this fixed point stable under local perturbations? This setting corresponds to dynamical decoding where local update rules permanently counteract to the corruption of memory.

Under certain assumptions for dimensional binary CA Toom's theorem applies and being an eroder is equivalent to the existence of stable fixed points. The most prominent example of such a CA is the North-East-Center (NEC) rule, also referred to as Toom's rule. It updates each cell by majority of the three (NEC) cells. For an initial i.i.d. configuration with probability $p$, the automaton calculates the majority if $p\neq1/2$, i.e. the automaton is an eroder. This corresponds to a decoder with threshold $p_\text{th}=1/2$. Considering the symmetry of the update rule and employing Toom's theorem it follows that the automaton must possess two robust fixed points. However, it is implausible, that the system can tolerate an error rate close to $1/2$ at every time step. This intuition can be made rigorous in the proof of phase coexistence. Consequently, the threshold for the fault tolerance of attractive states has to be lower then $1/2$. This lower threshold would correspond to the threshold of a dynamical decoder.

Using these insights into classical CA, we formulate the following rules of thumb for dynamical decoders: First, a dynamical decoder will usually also work as non-dynamical decoder. And second, the dynamical threshold is lower than the non-dynamical threshold. To avoid confusion, it should be stated that survival time simulations can also be performed with non-dynamic decoders. Here, both the decoding without measurement errors and the decoding in $2+1$ dimensions with measurement errors classify as non-dynamic. For such experiments the decoder threshold and the survival time threshold is identical.
 
\subsection{Semigroup representation}
An asynchronous cellular automaton can be understood as the evolution induced by a time independent Markovian master equation. At a constant rate, the coherent evolution is interrupted by quantum jumps corresponding to local unitary operations. Such a prescription precisely corresponds to a continuous one parameter semigroup, here acting on both the quantum state, as well as the effectively classical record of the anyon location and the classical auxiliary fields. The dynamical equation is given by 
\begin{equation}
	\frac{\partial \sigma }{\partial t} = \LLL( \sigma ) 
\end{equation}	
 where $\LLL$ is the Liouvillian decomposed as 
\begin{equation}
	\LLL:= \sum_j \LLL_j 
\end{equation}
and each local term takes the form 
\begin{equation}\LLL_j(\sigma)=\gamma_j ( L_j \sigma L_j^\dagger - L_j^\dagger L_j \sigma/2 -  \sigma L_j^\dagger L_j/2).
\end{equation} 
The $\{L_j\}$ are called Lindblad operators and $\{\gamma_j>0\}$ can be viewed as the rates of the dissipative processes. Specifically, we construct a Liouvillian where the geometrically local jump operators $L_j$ are constructed from $X_{p,\vec{v}}$, $m_{q,\vec{x}}$, $f_\vec{x}$, and $a_\vec{x}$, for all $\vec{x}$ and $\vec{v}$, with corresponding weights $\{\gamma_X,\gamma_m,\gamma_f,\gamma_a\}$. The $\gamma_f$ rates must now scale as $c\sim\log^2 L$ relative to all other rates to emulate the scaling of $c$ in the previous paragraph. In other words, asynchronous dynamical decoding can be understood as a  dissipative process 
\begin{equation}
\LLL = \LLL_{\CCC} + \LLL_\EEE,
\end{equation}
where the dissipative decoder reflected by $\LLL_{\CCC}$ continuously ``corrects'' an uncontrollable noise operator $\LLL_\EEE$ describing the actual noise. In this sense, the promise of ``fighting noise with noise'' is most manifestly realized: 
{Unwanted and equally uncontrollable noise processes are here corrected for by a controlled local dissipative process.}
We hence naturally bridge the gap between decoding and dissipative self-correcting \cite{PCC11,KWE13}. The Lindblad operators can be directly constructed from the local unitaries or stochastic maps of the update rules as follows: Consider a continuous-time classical stochastic process in which the unitary $O$ is applied to a quantum system at random times, uniformly distributed at a constant rate $\lambda>0$, such that the number of events in time interval $(t, t + \tau]$ 
follows a Poisson distribution with associated parameter $\lambda\tau$.
This classical stochastic process is on average on the level of the quantum system 
reflected by a quantum Markov process.

Similarly, the purely classical part can be embedded in a quantum system, the involved quantum states being diagonal. In this picture, there is a quantum mechanical 
equivalent $T_{\mathbf x}$ of an update rule 
$f_{\mathbf x}$ at a site ${\mathbf x}$
acting as 
\begin{equation}
	\rho\mapsto T_{\mathbf x} \rho T_{\mathbf x}^\dagger
\end{equation}	
on quantum states $\rho$.
Again, if the rule $f_{\mathbf x}$ is applied in a continuous asynchronous
fashion at rate $\lambda>0$, this is reflected by the quantum dynamical semi-group
of the above form with Lindblad operators $L_\mathbf x= T_{\mathbf x}$.
The overall asychronous cellular automaton 
can be embedded in a tri-partite quantum system with Hilbert space
${\cal H}\otimes {\cal L}\otimes {\cal V}$, with Lindblad operators $
\{L_j\}$ acting locally and at two tensor factors at the time. The actual
physical noise process is acting only on the first tensor factor 
${\cal H}$ associated with the physical quantum degrees of freedom.
 For a practical local dissipative decoder that does not have a threshold, see the interesting recent proposal in Ref.\ \cite{bardyn15}.

\subsection{System size dependencies}
The $\phi$-automaton decoder is characterized by the simplicity of its update rules and is particularly amenable to a representation as an asynchronous CA. However, there are some mild system size dependencies hidden in the construction that need to be accounted for. The first obvious one is the necessity to run the field updates for a number of rounds of order $\polylog L$ for every anyon move. Translated to the dissipative setting, this corresponds to the rates $\gamma_f$ scaling as $\polylog L$. In physical terms, this reflects the necessity of the field update operations to be increasingly fast with the size of the system. In practice, this is not much of an issue, as the field updates correspond to purely classical processing.

Secondly, in the rule for  anyon displacement, we need the field to have a resolution growing with the system size. If the field is encoded digitally, this corresponds to a local dimension of the CA lattice 
moderately scaling as $\log \log L$.  Finally, the anyon updates do not depend on the magnitude of the field, but rather move with unit probability, whenever there is a non-zero field gradient. This feature can also be adapted to the dissipative picture with a logarithmic overhead by digitally simulating a step function at the level of the jump operators. Thus, the CA decoder can naturally be represented as a local dissipative process, at the cost of merely logarithmically unbounded rates. \etc{These additional polylogarithmic overheads may be acceptable 
\jen{in many reasonable and realistic} physical set-ups, though \jen{conceptually speaking,} ultimately our work does not \jen{yet} deliver the widely sought after constant overhead dynamical decoder.}

\subsection{Dynamical decoders using directed signaling}
A dynamical CA decoder analyzed by Harrington avoids the required logarithmic speedup of our 3D decoder \cite{Har04,Michnicki}. The basic working principle is analogous to traditional MWPM heuristics \citep{Avi83}. The subsequent fusion of anyons over larger and larger distances is realized via \mjk{$O(\log L)$} hardware layers with each layer being responsible for a certain class of anyon distances. Most of the cells implement directed signaling between a few cells that coordinate the correction process. This approach requires inhomogeneous update rules, a certain structure inside the automaton and synchronous operation. The effective time scales are chosen such that the correction processes do not interfere between the layers. Ref.~\cite{Har04} provides numerics for {three lattice sizes that suggest} a threshold between $0.01\,\%$ and $0.1\,\%$. Ref.\ \cite{TerhalInPreparation} finds similar numbers for the crossings. 

It is worth noting that neither our cellular automaton decoders, nor the directed signaling ones in Refs.\ \cite{Har04,TerhalInPreparation} can conclusively claim numerical evidence for a fault-tolerant threshold. However, both have very similar behavior, with crossings shifting to lower values of $p$ as the system size increases, but providing very significant increases in survival time as a function of system size. The directed signaling decoder in Refs.\ \cite{Har04,TerhalInPreparation} are claimed to provably have a threshold for a steep hierarchical structure. We take this as encouraging evidence that our decoder will have a fault-tolerant threshold for low enough $p$.

\subsection{Existence of constant overhead decoders}
The fundamental challenge in building topological quantum memories is formulated in the established conjecture that self-correcting quantum memories only exist in four or higher 
dimensions {\cite{Yos11,PastawskiLimitations,HaahModules,Bre14}}. 
This \etc{leads} to the rule of thumb that 1D classical and 2D quantum memories share 
fundamental features and equally, 2D classical and 4D quantum memories. 
Classical memories in 2D can exhibit a thermal phase transition, {as, e.g., realized by a 
2D Ising model (for the connection between thermal phase transitions and quantum error correction,
see, e.g., Ref.\ \cite{LidarBrunQuantumErrorCorrection}).} 
Again in 2D, dynamical decoders for classical memories with constant overhead exist, {as realized by Toom's rule,
based on a two-dimensional binary cellular automaton
that is capable of calculating the majority in the system locally.}
The same applies to the 4D toric code \cite{DKL02} {for quantum error correction}, 
where the idea of using 
dynamical decoding for dissipative self correcting has already been studied \cite{PCC11}. While there are no thermal phase transitions in 1D for classical Hamiltonian systems, there do exist ordered phases in one dimensional interacting particle systems. This is constructively proven by the seminal work of G{\'a}cs \cite{Gac86,Gac01,Gra01}. In the light of dynamical decoders, G{\'a}cs' cellular automaton can be interpreted as an asynchronous dynamical decoder that removes string-like excitations from classical 1D systems using only constant overhead. Harrington has speculated that this ought to generalize to a two dimensional quantum code \cite{Har04}.  However, even in 1D, the G{\'a}cs algorithm is tremendously complex with no numerical or experimental realization known to the authors.  Analytic work estimates the G{\'a}cs 1D threshold at $\sim2^{-1000}$, so presently the outlook is poor for such an approach to yield a quantum decoder of practical relevance.\\

\section{Discussion and outlook} 
In this work we have introduced a class of dynamical decoders that are suited for operating on a large number of physical and logical qubits. Our proposal of a 3D $\phi$-automaton as an asynchronous dynamical decoder \mh{suggests} that very simple algorithms with modest overhead can exhibit significant error suppression, even in the presence of measurement errors. Also, we have shown how new decoding algorithms provide a blueprint for engineering  dissipative self-correcting memories. The arguments given in Ref.~\cite{LYP15} are likely to rule out the possibility to achi\mh{e}ve robustness in quantum memories by introducing additional terms in the Hamiltonian. 

Therefore, it seems imperative to circumvent no-go theorems for Hamiltonian self-correcting memories by either considering codes not covered by such theorems \cite{Bre14}; or to engineer dissipation that renders the code in effect self correcting. This constitutes an exciting perspective: 
Self-correcting quantum memories in 3D may well exist, if one lessens the requirement that the quantum
memories have to be of Hamiltonian nature and allows for Liouvillian dynamics.
Our 3D $\phi$-automaton decoder may not yet quite provide a final answer to those questions, since the corresponding Lindblad operators contain terms having a mild $L$ dependence scaling as $\log^2 L$. In general the design of dynamical decoders seems to inherit a trade-off between the degree of complexity in the local rules and the scaling of the local resources.

It provides a clear perspective, however, how such dissipative decoders can be designed.
The features of complete parallelization and asynchronous operation are not only relevant for the implementation also address the problem of wiring for on-chip architectures \cite{Altland}, 
and naturally incorporates asynchronous syndrome measurements, attractive for photonic chip based processors. Markovian dynamical decoders are therefore particularly interesting playground  for fault tolerance while providing intrinsic parallelization. 

\section{Acknowledgements}
We thank Ben Brown and Barbara Terhal for helpful discussions and for carefully reading a draft. 
JE is supported by the DFG (CRC 183), \jen{for which this work is specifically relevant for project B2, and}  (EI 519/7-1), the
ERC (TAQ), EU (AQuS), and the BMBF (Q.com).
ETC is supported by the EPSRC (grant EP/M024261/1). MJK is supported by the 
Carlsbergfond and the Villum fond. 


\begin{thebibliography}{47}%
\makeatletter
\providecommand \@ifxundefined [1]{%
 \@ifx{#1\undefined}
}%
\providecommand \@ifnum [1]{%
 \ifnum #1\expandafter \@firstoftwo
 \else \expandafter \@secondoftwo
 \fi
}%
\providecommand \@ifx [1]{%
 \ifx #1\expandafter \@firstoftwo
 \else \expandafter \@secondoftwo
 \fi
}%
\providecommand \natexlab [1]{#1}%
\providecommand \enquote  [1]{``#1''}%
\providecommand \bibnamefont  [1]{#1}%
\providecommand \bibfnamefont [1]{#1}%
\providecommand \citenamefont [1]{#1}%
\providecommand \href@noop [0]{\@secondoftwo}%
\providecommand \href [0]{\begingroup \@sanitize@url \@href}%
\providecommand \@href[1]{\@@startlink{#1}\@@href}%
\providecommand \@@href[1]{\endgroup#1\@@endlink}%
\providecommand \@sanitize@url [0]{\catcode `\\12\catcode `\$12\catcode
  `\&12\catcode `\#12\catcode `\^12\catcode `\_12\catcode `\%12\relax}%
\providecommand \@@startlink[1]{}%
\providecommand \@@endlink[0]{}%
\providecommand \url  [0]{\begingroup\@sanitize@url \@url }%
\providecommand \@url [1]{\endgroup\@href {#1}{\urlprefix }}%
\providecommand \urlprefix  [0]{URL }%
\providecommand \Eprint [0]{\href }%
\providecommand \doibase [0]{http://dx.doi.org/}%
\providecommand \selectlanguage [0]{\@gobble}%
\providecommand \bibinfo  [0]{\@secondoftwo}%
\providecommand \bibfield  [0]{\@secondoftwo}%
\providecommand \translation [1]{[#1]}%
\providecommand \BibitemOpen [0]{}%
\providecommand \bibitemStop [0]{}%
\providecommand \bibitemNoStop [0]{.\EOS\space}%
\providecommand \EOS [0]{\spacefactor3000\relax}%
\providecommand \BibitemShut  [1]{\csname bibitem#1\endcsname}%
\let\auto@bib@innerbib\@empty
\bibitem [{\citenamefont {Kitaev}(1997)}]{kitaev97}%
  \BibitemOpen
  \bibfield  {author} {\bibinfo {author} {\bibfnamefont {A.~Y.}\ \bibnamefont
  {Kitaev}},\ }in\ \href {\doibase 10.1007/b114679} {\emph {\bibinfo
  {booktitle} {Quantum communication, computing, and measurement}}}\ (\bibinfo
  {publisher} {Springer},\ \bibinfo {year} {1997})\ pp.\ \bibinfo {pages}
  {181--188}\BibitemShut {NoStop}%
\bibitem [{\citenamefont {{Dennis}}\ \emph {et~al.}(2002)\citenamefont
  {{Dennis}}, \citenamefont {{Kitaev}}, \citenamefont {{Landahl}},\ and\
  \citenamefont {{Preskill}}}]{DKL02}%
  \BibitemOpen
  \bibfield  {author} {\bibinfo {author} {\bibfnamefont {E.}~\bibnamefont
  {{Dennis}}}, \bibinfo {author} {\bibfnamefont {A.}~\bibnamefont {{Kitaev}}},
  \bibinfo {author} {\bibfnamefont {A.}~\bibnamefont {{Landahl}}}, \ and\
  \bibinfo {author} {\bibfnamefont {J.}~\bibnamefont {{Preskill}}},\ }\href
  {\doibase 10.1063/1.1499754} {\bibfield  {journal} {\bibinfo  {journal} {J.
  Math. Phys.}\ }\textbf {\bibinfo {volume} {43}},\ \bibinfo {pages} {4452}
  (\bibinfo {year} {2002})}\BibitemShut {NoStop}%
\bibitem [{\citenamefont {Nussinov}\ and\ \citenamefont
  {Ortiz}(2008)}]{Nussinov08}%
  \BibitemOpen
  \bibfield  {author} {\bibinfo {author} {\bibfnamefont {Z.}~\bibnamefont
  {Nussinov}}\ and\ \bibinfo {author} {\bibfnamefont {G.}~\bibnamefont
  {Ortiz}},\ }\href {\doibase 10.1103/PhysRevB.77.064302} {\bibfield  {journal}
  {\bibinfo  {journal} {Phys. Rev. B}\ }\textbf {\bibinfo {volume} {77}},\
  \bibinfo {pages} {064302} (\bibinfo {year} {2008})}\BibitemShut {NoStop}%
\bibitem [{\citenamefont {{Hamma}}\ \emph {et~al.}(2009)\citenamefont
  {{Hamma}}, \citenamefont {{Castelnovo}},\ and\ \citenamefont
  {{Chamon}}}]{HCC09}%
  \BibitemOpen
  \bibfield  {author} {\bibinfo {author} {\bibfnamefont {A.}~\bibnamefont
  {{Hamma}}}, \bibinfo {author} {\bibfnamefont {C.}~\bibnamefont
  {{Castelnovo}}}, \ and\ \bibinfo {author} {\bibfnamefont {C.}~\bibnamefont
  {{Chamon}}},\ }\href {\doibase 10.1103/PhysRevB.79.245122} {\bibfield
  {journal} {\bibinfo  {journal} {Phys. Rev. B}\ }\textbf {\bibinfo {volume}
  {79}},\ \bibinfo {pages} {245122} (\bibinfo {year} {2009})}\BibitemShut
  {NoStop}%
\bibitem [{\citenamefont {Alicki}\ \emph {et~al.}(2009)\citenamefont {Alicki},
  \citenamefont {Fannes},\ and\ \citenamefont {Horodecki}}]{AFH09}%
  \BibitemOpen
  \bibfield  {author} {\bibinfo {author} {\bibfnamefont {R.}~\bibnamefont
  {Alicki}}, \bibinfo {author} {\bibfnamefont {M.}~\bibnamefont {Fannes}}, \
  and\ \bibinfo {author} {\bibfnamefont {M.}~\bibnamefont {Horodecki}},\ }\href
  {http://stacks.iop.org/1751-8121/42/i=6/a=065303} {\bibfield  {journal}
  {\bibinfo  {journal} {J. Phys. A}\ }\textbf {\bibinfo {volume} {42}},\
  \bibinfo {pages} {065303} (\bibinfo {year} {2009})}\BibitemShut {NoStop}%
\bibitem [{\citenamefont {{Landon-Cardinal}}\ \emph {et~al.}(2015)\citenamefont
  {{Landon-Cardinal}}, \citenamefont {{Yoshida}}, \citenamefont {{Poulin}},\
  and\ \citenamefont {{Preskill}}}]{LYP15}%
  \BibitemOpen
  \bibfield  {author} {\bibinfo {author} {\bibfnamefont {O.}~\bibnamefont
  {{Landon-Cardinal}}}, \bibinfo {author} {\bibfnamefont {B.}~\bibnamefont
  {{Yoshida}}}, \bibinfo {author} {\bibfnamefont {D.}~\bibnamefont {{Poulin}}},
  \ and\ \bibinfo {author} {\bibfnamefont {J.}~\bibnamefont {{Preskill}}},\
  }\href {\doibase 10.1103/PhysRevA.91.032303} {\bibfield  {journal} {\bibinfo
  {journal} {Phys. Rev. A}\ }\textbf {\bibinfo {volume} {91}},\ \bibinfo
  {pages} {032303} (\bibinfo {year} {2015})}\BibitemShut {NoStop}%
\bibitem [{\citenamefont {Bardyn}\ and\ \citenamefont
  {Karzig}(2016)}]{bardyn15}%
  \BibitemOpen
  \bibfield  {author} {\bibinfo {author} {\bibfnamefont {C.-E.}\ \bibnamefont
  {Bardyn}}\ and\ \bibinfo {author} {\bibfnamefont {T.}~\bibnamefont
  {Karzig}},\ }\href@noop {} {\bibfield  {journal} {\bibinfo  {journal} {Phys.
  Rev. B}\ }\textbf {\bibinfo {volume} {94}},\ \bibinfo {pages} {094303}
  (\bibinfo {year} {2016})}\BibitemShut {NoStop}%
\bibitem [{\citenamefont {Taminiau}\ \emph {et~al.}(2014)\citenamefont
  {Taminiau}, \citenamefont {Cramer}, \citenamefont {van~der Sar},
  \citenamefont {Dobrovitski},\ and\ \citenamefont {Hanson}}]{Hanson14}%
  \BibitemOpen
  \bibfield  {author} {\bibinfo {author} {\bibfnamefont {T.~H.}\ \bibnamefont
  {Taminiau}}, \bibinfo {author} {\bibfnamefont {J.}~\bibnamefont {Cramer}},
  \bibinfo {author} {\bibfnamefont {T.}~\bibnamefont {van~der Sar}}, \bibinfo
  {author} {\bibfnamefont {V.~V.}\ \bibnamefont {Dobrovitski}}, \ and\ \bibinfo
  {author} {\bibfnamefont {R.}~\bibnamefont {Hanson}},\ }\href
  {http://dx.doi.org/10.1038/nnano.2014.2} {\bibfield  {journal} {\bibinfo
  {journal} {Nature Nano.}\ }\textbf {\bibinfo {volume} {9}},\ \bibinfo {pages}
  {171} (\bibinfo {year} {2014})}\BibitemShut {NoStop}%
\bibitem [{\citenamefont {Waldherr}\ \emph {et~al.}(2014)\citenamefont
  {Waldherr}, \citenamefont {Wang}, \citenamefont {Zaiser}, \citenamefont
  {Jamali}, \citenamefont {Schulte-Herbruggen}, \citenamefont {Abe},
  \citenamefont {Ohshima}, \citenamefont {Isoya}, \citenamefont {Du},
  \citenamefont {Neumann},\ and\ \citenamefont {Wrachtrup}}]{Wrachtrup14}%
  \BibitemOpen
  \bibfield  {author} {\bibinfo {author} {\bibfnamefont {G.}~\bibnamefont
  {Waldherr}}, \bibinfo {author} {\bibfnamefont {Y.}~\bibnamefont {Wang}},
  \bibinfo {author} {\bibfnamefont {S.}~\bibnamefont {Zaiser}}, \bibinfo
  {author} {\bibfnamefont {M.}~\bibnamefont {Jamali}}, \bibinfo {author}
  {\bibfnamefont {T.}~\bibnamefont {Schulte-Herbruggen}}, \bibinfo {author}
  {\bibfnamefont {H.}~\bibnamefont {Abe}}, \bibinfo {author} {\bibfnamefont
  {T.}~\bibnamefont {Ohshima}}, \bibinfo {author} {\bibfnamefont
  {J.}~\bibnamefont {Isoya}}, \bibinfo {author} {\bibfnamefont {J.~F.}\
  \bibnamefont {Du}}, \bibinfo {author} {\bibfnamefont {P.}~\bibnamefont
  {Neumann}}, \ and\ \bibinfo {author} {\bibfnamefont {J.}~\bibnamefont
  {Wrachtrup}},\ }\href {http://dx.doi.org/10.1038/nature12919} {\bibfield
  {journal} {\bibinfo  {journal} {Nature}\ }\textbf {\bibinfo {volume} {506}},\
  \bibinfo {pages} {204} (\bibinfo {year} {2014})}\BibitemShut {NoStop}%
\bibitem [{\citenamefont {Sun}\ \emph {et~al.}(2014)\citenamefont {Sun},
  \citenamefont {Petrenko}, \citenamefont {Leghtas}, \citenamefont {Vlastakis},
  \citenamefont {Kirchmair}, \citenamefont {Sliwa}, \citenamefont {Narla},
  \citenamefont {Hatridge}, \citenamefont {Shankar}, \citenamefont {Blumoff},
  \citenamefont {Frunzio}, \citenamefont {Mirrahimi}, \citenamefont {Devoret},\
  and\ \citenamefont {Schoelkopf}}]{Sun14}%
  \BibitemOpen
  \bibfield  {author} {\bibinfo {author} {\bibfnamefont {L.}~\bibnamefont
  {Sun}}, \bibinfo {author} {\bibfnamefont {A.}~\bibnamefont {Petrenko}},
  \bibinfo {author} {\bibfnamefont {Z.}~\bibnamefont {Leghtas}}, \bibinfo
  {author} {\bibfnamefont {B.}~\bibnamefont {Vlastakis}}, \bibinfo {author}
  {\bibfnamefont {G.}~\bibnamefont {Kirchmair}}, \bibinfo {author}
  {\bibfnamefont {K.~M.}\ \bibnamefont {Sliwa}}, \bibinfo {author}
  {\bibfnamefont {A.}~\bibnamefont {Narla}}, \bibinfo {author} {\bibfnamefont
  {M.}~\bibnamefont {Hatridge}}, \bibinfo {author} {\bibfnamefont
  {S.}~\bibnamefont {Shankar}}, \bibinfo {author} {\bibfnamefont
  {J.}~\bibnamefont {Blumoff}}, \bibinfo {author} {\bibfnamefont
  {L.}~\bibnamefont {Frunzio}}, \bibinfo {author} {\bibfnamefont
  {M.}~\bibnamefont {Mirrahimi}}, \bibinfo {author} {\bibfnamefont {M.~H.}\
  \bibnamefont {Devoret}}, \ and\ \bibinfo {author} {\bibfnamefont {R.~J.}\
  \bibnamefont {Schoelkopf}},\ }\href {http://dx.doi.org/10.1038/nature13436}
  {\bibfield  {journal} {\bibinfo  {journal} {Nature}\ }\textbf {\bibinfo
  {volume} {511}},\ \bibinfo {pages} {444} (\bibinfo {year}
  {2014})}\BibitemShut {NoStop}%
\bibitem [{\citenamefont {Kelly}\ \emph {et~al.}(2015)\citenamefont {Kelly},
  \citenamefont {Barends}, \citenamefont {Fowler}, \citenamefont {Megrant},
  \citenamefont {Jeffrey}, \citenamefont {White}, \citenamefont {Sank},
  \citenamefont {Mutus}, \citenamefont {Campbell}, \citenamefont {Chen},
  \citenamefont {Chen}, \citenamefont {Chiaro}, \citenamefont {Dunsworth},
  \citenamefont {Hoi}, \citenamefont {Neill}, \citenamefont {O'Malley},
  \citenamefont {Quintana}, \citenamefont {Roushan}, \citenamefont
  {Vainsencher}, \citenamefont {Wenner}, \citenamefont {Cleland},\ and\
  \citenamefont {Martinis}}]{Martinis15}%
  \BibitemOpen
  \bibfield  {author} {\bibinfo {author} {\bibfnamefont {J.}~\bibnamefont
  {Kelly}}, \bibinfo {author} {\bibfnamefont {R.}~\bibnamefont {Barends}},
  \bibinfo {author} {\bibfnamefont {A.~G.}\ \bibnamefont {Fowler}}, \bibinfo
  {author} {\bibfnamefont {A.}~\bibnamefont {Megrant}}, \bibinfo {author}
  {\bibfnamefont {E.}~\bibnamefont {Jeffrey}}, \bibinfo {author} {\bibfnamefont
  {T.~C.}\ \bibnamefont {White}}, \bibinfo {author} {\bibfnamefont
  {D.}~\bibnamefont {Sank}}, \bibinfo {author} {\bibfnamefont {J.~Y.}\
  \bibnamefont {Mutus}}, \bibinfo {author} {\bibfnamefont {B.}~\bibnamefont
  {Campbell}}, \bibinfo {author} {\bibfnamefont {Y.}~\bibnamefont {Chen}},
  \bibinfo {author} {\bibfnamefont {Z.}~\bibnamefont {Chen}}, \bibinfo {author}
  {\bibfnamefont {B.}~\bibnamefont {Chiaro}}, \bibinfo {author} {\bibfnamefont
  {A.}~\bibnamefont {Dunsworth}}, \bibinfo {author} {\bibfnamefont {I.~C.}\
  \bibnamefont {Hoi}}, \bibinfo {author} {\bibfnamefont {C.}~\bibnamefont
  {Neill}}, \bibinfo {author} {\bibfnamefont {P.~J.~J.}\ \bibnamefont
  {O'Malley}}, \bibinfo {author} {\bibfnamefont {C.}~\bibnamefont {Quintana}},
  \bibinfo {author} {\bibfnamefont {P.}~\bibnamefont {Roushan}}, \bibinfo
  {author} {\bibfnamefont {A.}~\bibnamefont {Vainsencher}}, \bibinfo {author}
  {\bibfnamefont {J.}~\bibnamefont {Wenner}}, \bibinfo {author} {\bibfnamefont
  {A.~N.}\ \bibnamefont {Cleland}}, \ and\ \bibinfo {author} {\bibfnamefont
  {J.~M.}\ \bibnamefont {Martinis}},\ }\href
  {http://dx.doi.org/10.1038/nature14270} {\bibfield  {journal} {\bibinfo
  {journal} {Nature}\ }\textbf {\bibinfo {volume} {519}},\ \bibinfo {pages}
  {66} (\bibinfo {year} {2015})}\BibitemShut {NoStop}%
\bibitem [{\citenamefont {Terhal}(2015)}]{TerhalReview}%
  \BibitemOpen
  \bibfield  {author} {\bibinfo {author} {\bibfnamefont {B.}~\bibnamefont
  {Terhal}},\ }\href {\doibase http://dx.doi.org/10.1103/RevModPhys.87.307}
  {\bibfield  {journal} {\bibinfo  {journal} {Rev. Mod. Phys.}\ }\textbf
  {\bibinfo {volume} {87}},\ \bibinfo {pages} {307} (\bibinfo {year}
  {2015})}\BibitemShut {NoStop}%
\bibitem [{\citenamefont {{Wang}}\ \emph {et~al.}(2003)\citenamefont {{Wang}},
  \citenamefont {{Harrington}},\ and\ \citenamefont {{Preskill}}}]{WHP03}%
  \BibitemOpen
  \bibfield  {author} {\bibinfo {author} {\bibfnamefont {C.}~\bibnamefont
  {{Wang}}}, \bibinfo {author} {\bibfnamefont {J.}~\bibnamefont
  {{Harrington}}}, \ and\ \bibinfo {author} {\bibfnamefont {J.}~\bibnamefont
  {{Preskill}}},\ }\href {\doibase 10.1016/S0003-4916(02)00019-2} {\bibfield
  {journal} {\bibinfo  {journal} {Ann. Phys.}\ }\textbf {\bibinfo {volume}
  {303}},\ \bibinfo {pages} {31} (\bibinfo {year} {2003})}\BibitemShut
  {NoStop}%
\bibitem [{\citenamefont {{Fowler}}\ \emph {et~al.}(2012)\citenamefont
  {{Fowler}}, \citenamefont {{Whiteside}},\ and\ \citenamefont
  {{Hollenberg}}}]{FWH12b}%
  \BibitemOpen
  \bibfield  {author} {\bibinfo {author} {\bibfnamefont {A.~G.}\ \bibnamefont
  {{Fowler}}}, \bibinfo {author} {\bibfnamefont {A.~C.}\ \bibnamefont
  {{Whiteside}}}, \ and\ \bibinfo {author} {\bibfnamefont {L.~C.~L.}\
  \bibnamefont {{Hollenberg}}},\ }\href {\doibase 10.1103/PhysRevA.86.042313}
  {\bibfield  {journal} {\bibinfo  {journal} {Phys. Rev. A}\ }\textbf {\bibinfo
  {volume} {86}},\ \bibinfo {pages} {042313} (\bibinfo {year}
  {2012})}\BibitemShut {NoStop}%
\bibitem [{\citenamefont {{Duclos-Cianci}}\ and\ \citenamefont
  {{Poulin}}(2010)}]{DP10b}%
  \BibitemOpen
  \bibfield  {author} {\bibinfo {author} {\bibfnamefont {G.}~\bibnamefont
  {{Duclos-Cianci}}}\ and\ \bibinfo {author} {\bibfnamefont {D.}~\bibnamefont
  {{Poulin}}},\ }\href {\doibase 10.1103/PhysRevLett.104.050504} {\bibfield
  {journal} {\bibinfo  {journal} {Phys. Rev. Lett.}\ }\textbf {\bibinfo
  {volume} {104}},\ \bibinfo {pages} {050504} (\bibinfo {year}
  {2010})}\BibitemShut {NoStop}%
\bibitem [{\citenamefont {{Bravyi}}\ and\ \citenamefont {{Haah}}(2013)}]{BH13}%
  \BibitemOpen
  \bibfield  {author} {\bibinfo {author} {\bibfnamefont {S.}~\bibnamefont
  {{Bravyi}}}\ and\ \bibinfo {author} {\bibfnamefont {J.}~\bibnamefont
  {{Haah}}},\ }\href {\doibase 10.1103/PhysRevLett.111.200501} {\bibfield
  {journal} {\bibinfo  {journal} {Phys. Rev. Lett.}\ }\textbf {\bibinfo
  {volume} {111}},\ \bibinfo {pages} {200501} (\bibinfo {year}
  {2013})}\BibitemShut {NoStop}%
\bibitem [{\citenamefont {Wootton}\ and\ \citenamefont {Loss}(2012)}]{WL12}%
  \BibitemOpen
  \bibfield  {author} {\bibinfo {author} {\bibfnamefont {J.~R.}\ \bibnamefont
  {Wootton}}\ and\ \bibinfo {author} {\bibfnamefont {D.}~\bibnamefont {Loss}},\
  }\href {\doibase 10.1103/PhysRevLett.109.160503} {\bibfield  {journal}
  {\bibinfo  {journal} {Phys. Rev. Lett.}\ }\textbf {\bibinfo {volume} {109}},\
  \bibinfo {pages} {160503} (\bibinfo {year} {2012})}\BibitemShut {NoStop}%
\bibitem [{\citenamefont {Hutter}\ \emph {et~al.}(2014)\citenamefont {Hutter},
  \citenamefont {Wootton},\ and\ \citenamefont {Loss}}]{HWL14}%
  \BibitemOpen
  \bibfield  {author} {\bibinfo {author} {\bibfnamefont {A.}~\bibnamefont
  {Hutter}}, \bibinfo {author} {\bibfnamefont {J.~R.}\ \bibnamefont {Wootton}},
  \ and\ \bibinfo {author} {\bibfnamefont {D.}~\bibnamefont {Loss}},\ }\href
  {\doibase 10.1103/PhysRevA.89.022326} {\bibfield  {journal} {\bibinfo
  {journal} {Phys. Rev. A}\ }\textbf {\bibinfo {volume} {89}},\ \bibinfo
  {pages} {022326} (\bibinfo {year} {2014})}\BibitemShut {NoStop}%
\bibitem [{\citenamefont {Herold}\ \emph {et~al.}(2015)\citenamefont {Herold},
  \citenamefont {Campbell}, \citenamefont {Eisert},\ and\ \citenamefont
  {Kastoryano}}]{HCEK14}%
  \BibitemOpen
  \bibfield  {author} {\bibinfo {author} {\bibfnamefont {M.}~\bibnamefont
  {Herold}}, \bibinfo {author} {\bibfnamefont {E.~T.}\ \bibnamefont
  {Campbell}}, \bibinfo {author} {\bibfnamefont {J.}~\bibnamefont {Eisert}}, \
  and\ \bibinfo {author} {\bibfnamefont {M.~J.}\ \bibnamefont {Kastoryano}},\
  }\href {\doibase 10.1038/npjqi.2015.10} {\bibfield  {journal} {\bibinfo
  {journal} {npj Quant. Inf.}\ }\textbf {\bibinfo {volume} {1}}, 15010
  (\bibinfo {year} {2015}),\ 10.1038/npjqi.2015.10}\BibitemShut {NoStop}%
\bibitem [{\citenamefont {Harrington}(2004)}]{Har04}%
  \BibitemOpen
  \bibfield  {author} {\bibinfo {author} {\bibfnamefont {J.~W.}\ \bibnamefont
  {Harrington}},\ }\emph {\bibinfo {title} {Analysis of quantum
  error-correcting codes: symplectic lattice codes and toric codes}},\
  \href@noop {} {Ph.D. thesis} (\bibinfo {year} {2004}),\ \bibinfo {note}
  {\url{http://thesis.library.caltech.edu/1747/1/jimh_thesis.pdf}}\BibitemShut
  {NoStop}%
\bibitem [{Note1()}]{Note1}%
  \BibitemOpen
  \bibinfo {note} {\protect \emph {Private communication}, B.\ Terhal
  (2016).}\BibitemShut {Stop}%
\bibitem [{\citenamefont {Hutter}\ \emph {et~al.}(2015)\citenamefont {Hutter},
  \citenamefont {Loss},\ and\ \citenamefont {Wootton}}]{HutterHDRG}%
  \BibitemOpen
  \bibfield  {author} {\bibinfo {author} {\bibfnamefont {A.}~\bibnamefont
  {Hutter}}, \bibinfo {author} {\bibfnamefont {D.}~\bibnamefont {Loss}}, \ and\
  \bibinfo {author} {\bibfnamefont {J.~R.}\ \bibnamefont {Wootton}},\ }\href
  {http://stacks.iop.org/1367-2630/17/i=3/a=035017} {\bibfield  {journal}
  {\bibinfo  {journal} {New J. Phys.}\ }\textbf {\bibinfo {volume} {17}},\
  \bibinfo {pages} {035017} (\bibinfo {year} {2015})}\BibitemShut {NoStop}%
\bibitem [{\citenamefont {Dauphinais}\ and\ \citenamefont
  {Poulin}(2016)}]{PoulinAnyons}%
  \BibitemOpen
  \bibfield  {author} {\bibinfo {author} {\bibfnamefont {G.}~\bibnamefont
  {Dauphinais}}\ and\ \bibinfo {author} {\bibfnamefont {D.}~\bibnamefont
  {Poulin}},\ }\href@noop {} {\bibfield  {journal} {\bibinfo  {journal} {arXiv
  preprint arXiv:1607.02159}\ } (\bibinfo {year} {2016})}\BibitemShut {NoStop}%
\bibitem [{\citenamefont {Cabrillo}\ \emph {et~al.}(1999)\citenamefont
  {Cabrillo}, \citenamefont {Cirac}, \citenamefont {Garcia-Fernandez},\ and\
  \citenamefont {Zoller}}]{CCGZ99}%
  \BibitemOpen
  \bibfield  {author} {\bibinfo {author} {\bibfnamefont {C.}~\bibnamefont
  {Cabrillo}}, \bibinfo {author} {\bibfnamefont {J.~I.}\ \bibnamefont {Cirac}},
  \bibinfo {author} {\bibfnamefont {P.}~\bibnamefont {Garcia-Fernandez}}, \
  and\ \bibinfo {author} {\bibfnamefont {P.}~\bibnamefont {Zoller}},\ }\href
  {\doibase http://dx.doi.org/10.1103/PhysRevA.59.1025} {\bibfield  {journal}
  {\bibinfo  {journal} {Phys. Rev. A}\ }\textbf {\bibinfo {volume} {59}},\
  \bibinfo {pages} {1025} (\bibinfo {year} {1999})}\BibitemShut {NoStop}%
\bibitem [{\citenamefont {Lim}\ \emph {et~al.}(2005)\citenamefont {Lim},
  \citenamefont {Beige},\ and\ \citenamefont {Kwek}}]{LBK05}%
  \BibitemOpen
  \bibfield  {author} {\bibinfo {author} {\bibfnamefont {Y.~L.}\ \bibnamefont
  {Lim}}, \bibinfo {author} {\bibfnamefont {A.}~\bibnamefont {Beige}}, \ and\
  \bibinfo {author} {\bibfnamefont {L.~C.}\ \bibnamefont {Kwek}},\ }\href
  {\doibase http://dx.doi.org/10.1103/PhysRevLett.95.030505} {\bibfield
  {journal} {\bibinfo  {journal} {Phys. Rev. Lett.}\ }\textbf {\bibinfo
  {volume} {95}},\ \bibinfo {pages} {030505} (\bibinfo {year}
  {2005})}\BibitemShut {NoStop}%
\bibitem [{\citenamefont {Barrett}\ and\ \citenamefont {Kok}(2005)}]{BK05}%
  \BibitemOpen
  \bibfield  {author} {\bibinfo {author} {\bibfnamefont {S.~D.}\ \bibnamefont
  {Barrett}}\ and\ \bibinfo {author} {\bibfnamefont {P.}~\bibnamefont {Kok}},\
  }\href {\doibase http://dx.doi.org/10.1103/PhysRevA.71.060310} {\bibfield
  {journal} {\bibinfo  {journal} {Phys. Rev. A}\ }\textbf {\bibinfo {volume}
  {71}},\ \bibinfo {pages} {060310} (\bibinfo {year} {2005})}\BibitemShut
  {NoStop}%
\bibitem [{\citenamefont {Campbell}\ and\ \citenamefont
  {Benjamin}(2008)}]{CB08}%
  \BibitemOpen
  \bibfield  {author} {\bibinfo {author} {\bibfnamefont {E.~T.}\ \bibnamefont
  {Campbell}}\ and\ \bibinfo {author} {\bibfnamefont {S.~C.}\ \bibnamefont
  {Benjamin}},\ }\href {\doibase
  http://dx.doi.org/10.1103/PhysRevLett.101.130502} {\bibfield  {journal}
  {\bibinfo  {journal} {Phys. Rev. Lett.}\ }\textbf {\bibinfo {volume} {101}},\
  \bibinfo {pages} {130502} (\bibinfo {year} {2008})}\BibitemShut {NoStop}%
\bibitem [{\citenamefont {Nickerson}\ \emph {et~al.}(2013)\citenamefont
  {Nickerson}, \citenamefont {Li},\ and\ \citenamefont {Benjamin}}]{NYB13}%
  \BibitemOpen
  \bibfield  {author} {\bibinfo {author} {\bibfnamefont {N.~H.}\ \bibnamefont
  {Nickerson}}, \bibinfo {author} {\bibfnamefont {Y.}~\bibnamefont {Li}}, \
  and\ \bibinfo {author} {\bibfnamefont {S.~C.}\ \bibnamefont {Benjamin}},\
  }\href {\doibase http://dx.doi.org/10.1038/ncomms2773} {\bibfield  {journal}
  {\bibinfo  {journal} {Nature Comm.}\ }\textbf {\bibinfo {volume} {4}},\
  \bibinfo {pages} {1756} (\bibinfo {year} {2013})}\BibitemShut {NoStop}%
\bibitem [{\citenamefont {Terhal}\ \emph {et~al.}(2012)\citenamefont {Terhal},
  \citenamefont {Hassler},\ and\ \citenamefont
  {DiVincenzo}}]{TerhalDiVincenzo}%
  \BibitemOpen
  \bibfield  {author} {\bibinfo {author} {\bibfnamefont {B.~M.}\ \bibnamefont
  {Terhal}}, \bibinfo {author} {\bibfnamefont {F.}~\bibnamefont {Hassler}}, \
  and\ \bibinfo {author} {\bibfnamefont {D.~P.}\ \bibnamefont {DiVincenzo}},\
  }\href {\doibase 10.1103/PhysRevLett.108.260504} {\bibfield  {journal}
  {\bibinfo  {journal} {Phys. Rev. Lett.}\ }\textbf {\bibinfo {volume} {108}},\
  \bibinfo {pages} {260504} (\bibinfo {year} {2012})}\BibitemShut {NoStop}%
\bibitem [{\citenamefont {Landau}\ \emph {et~al.}(2016)\citenamefont {Landau},
  \citenamefont {Plugge}, \citenamefont {Sela}, \citenamefont {Altland},
  \citenamefont {Albrecht},\ and\ \citenamefont {Egger}}]{Altland}%
  \BibitemOpen
  \bibfield  {author} {\bibinfo {author} {\bibfnamefont {L.~A.}\ \bibnamefont
  {Landau}}, \bibinfo {author} {\bibfnamefont {S.}~\bibnamefont {Plugge}},
  \bibinfo {author} {\bibfnamefont {E.}~\bibnamefont {Sela}}, \bibinfo {author}
  {\bibfnamefont {A.}~\bibnamefont {Altland}}, \bibinfo {author} {\bibfnamefont
  {S.~M.}\ \bibnamefont {Albrecht}}, \ and\ \bibinfo {author} {\bibfnamefont
  {R.}~\bibnamefont {Egger}},\ }\href {\doibase 10.1103/PhysRevLett.116.050501}
  {\bibfield  {journal} {\bibinfo  {journal} {Phys. Rev. Lett.}\ }\textbf
  {\bibinfo {volume} {116}},\ \bibinfo {pages} {050501} (\bibinfo {year}
  {2016})}\BibitemShut {NoStop}%
\bibitem [{\citenamefont {Breuckmann}\ \emph {et~al.}(2016)\citenamefont
  {Breuckmann}, \citenamefont {Michels}, \citenamefont {Duivenvoorden},\ and\
  \citenamefont {Terhal}}]{TerhalInPreparation}%
  \BibitemOpen
  \bibfield  {author} {\bibinfo {author} {\bibfnamefont {N.~P.}\ \bibnamefont
  {Breuckmann}}, \bibinfo {author} {\bibfnamefont {D.}~\bibnamefont {Michels}},
  \bibinfo {author} {\bibfnamefont {K.}~\bibnamefont {Duivenvoorden}}, \ and\
  \bibinfo {author} {\bibfnamefont {B.~M.}\ \bibnamefont {Terhal}},\
  }\href@noop {} {\enquote {\bibinfo {title} {Local decoders for the 2d and 4d
  toric code},}\ } (\bibinfo {year} {2016}),\ \bibinfo {note}
  {arXiv:1609.00510}\BibitemShut {NoStop}%
\bibitem [{\citenamefont {{Pedrocchi}}\ \emph {et~al.}(2011)\citenamefont
  {{Pedrocchi}}, \citenamefont {{Chesi}},\ and\ \citenamefont
  {{Loss}}}]{PCL11}%
  \BibitemOpen
  \bibfield  {author} {\bibinfo {author} {\bibfnamefont {F.~L.}\ \bibnamefont
  {{Pedrocchi}}}, \bibinfo {author} {\bibfnamefont {S.}~\bibnamefont
  {{Chesi}}}, \ and\ \bibinfo {author} {\bibfnamefont {D.}~\bibnamefont
  {{Loss}}},\ }\href {\doibase 10.1103/PhysRevB.83.115415} {\bibfield
  {journal} {\bibinfo  {journal} {Phys. Rev. B}\ }\textbf {\bibinfo {volume}
  {83}},\ \bibinfo {pages} {115415} (\bibinfo {year} {2011})},\ \Eprint
  {http://arxiv.org/abs/1011.3762} {arXiv:1011.3762} \BibitemShut {NoStop}%
\bibitem [{\citenamefont {{Pedrocchi}}\ \emph {et~al.}(2013)\citenamefont
  {{Pedrocchi}}, \citenamefont {{Hutter}}, \citenamefont {{Wootton}},\ and\
  \citenamefont {{Loss}}}]{PHW13}%
  \BibitemOpen
  \bibfield  {author} {\bibinfo {author} {\bibfnamefont {F.}~\bibnamefont
  {{Pedrocchi}}}, \bibinfo {author} {\bibfnamefont {A.}~\bibnamefont
  {{Hutter}}}, \bibinfo {author} {\bibfnamefont {J.}~\bibnamefont {{Wootton}}},
  \ and\ \bibinfo {author} {\bibfnamefont {D.}~\bibnamefont {{Loss}}},\ }\href
  {\doibase 10.1103/PhysRevA.88.062313} {\bibfield  {journal} {\bibinfo
  {journal} {Phys. Rev. A}\ }\textbf {\bibinfo {volume} {88}},\ \bibinfo
  {pages} {062313} (\bibinfo {year} {2013})} \BibitemShut {NoStop}%
\bibitem [{\citenamefont {Bomb\'{\i}n}(2015)}]{Bom14}%
  \BibitemOpen
  \bibfield  {author} {\bibinfo {author} {\bibfnamefont {H.}~\bibnamefont
  {Bomb\'{\i}n}},\ }\href {\doibase 10.1103/PhysRevX.5.031043} {\bibfield
  {journal} {\bibinfo  {journal} {Phys. Rev. X}\ }\textbf {\bibinfo {volume}
  {5}},\ \bibinfo {pages} {031043} (\bibinfo {year} {2015})}\BibitemShut
  {NoStop}%
\bibitem [{\citenamefont {Brown}\ \emph {et~al.}(2016)\citenamefont {Brown},
  \citenamefont {Nickerson},\ and\ \citenamefont {Browne}}]{BNB15}%
  \BibitemOpen
  \bibfield  {author} {\bibinfo {author} {\bibfnamefont {B.~J.}\ \bibnamefont
  {Brown}}, \bibinfo {author} {\bibfnamefont {N.~H.}\ \bibnamefont
  {Nickerson}}, \ and\ \bibinfo {author} {\bibfnamefont {D.~E.}\ \bibnamefont
  {Browne}},\ }\href {\doibase doi:10.1038/ncomms12302} {\bibfield  {journal}
  {\bibinfo  {journal} {Nature Comm.}\ }\textbf {\bibinfo {volume} {7}},\
  \bibinfo {pages} {12302} (\bibinfo {year} {2016})}\BibitemShut {NoStop}%
\bibitem [{\citenamefont {{G{\'a}cs}}(1986)}]{Gac86}%
  \BibitemOpen
  \bibfield  {author} {\bibinfo {author} {\bibfnamefont {P.}~\bibnamefont
  {{G{\'a}cs}}},\ }\href {\doibase 10.1016/0022-0000(86)90002-4} {\bibfield
  {journal} {\bibinfo  {journal} {J. Comp. Sys. Sc.}\ }\textbf {\bibinfo
  {volume} {32}},\ \bibinfo {pages} {15} (\bibinfo {year} {1986})}\BibitemShut
  {NoStop}%
\bibitem [{\citenamefont {{G{\'a}cs}}(2001)}]{Gac01}%
  \BibitemOpen
  \bibfield  {author} {\bibinfo {author} {\bibfnamefont {P.}~\bibnamefont
  {{G{\'a}cs}}},\ }\href {\doibase 10.1023/A:1004823720305} {\bibfield
  {journal} {\bibinfo  {journal} {J. Stat. Phys.}\ }\textbf {\bibinfo {volume}
  {103}},\ \bibinfo {pages} {45} (\bibinfo {year} {2001})}\BibitemShut
  {NoStop}%
\bibitem [{\citenamefont {{Gray}}(2001)}]{Gra01}%
  \BibitemOpen
  \bibfield  {author} {\bibinfo {author} {\bibfnamefont {L.~F.}\ \bibnamefont
  {{Gray}}},\ }\href {\doibase 10.1023/A:1004824203467} {\bibfield  {journal}
  {\bibinfo  {journal} {J. Stat. Phys.}\ }\textbf {\bibinfo {volume} {103}},\
  \bibinfo {pages} {1} (\bibinfo {year} {2001})}\BibitemShut {NoStop}%
\bibitem [{\citenamefont {{Pastawski}}\ \emph {et~al.}(2011)\citenamefont
  {{Pastawski}}, \citenamefont {{Clemente}},\ and\ \citenamefont
  {{Cirac}}}]{PCC11}%
  \BibitemOpen
  \bibfield  {author} {\bibinfo {author} {\bibfnamefont {F.}~\bibnamefont
  {{Pastawski}}}, \bibinfo {author} {\bibfnamefont {L.}~\bibnamefont
  {{Clemente}}}, \ and\ \bibinfo {author} {\bibfnamefont {J.~I.}\ \bibnamefont
  {{Cirac}}},\ }\href {\doibase 10.1103/PhysRevA.83.012304} {\bibfield
  {journal} {\bibinfo  {journal} {Phys. Rev. A}\ }\textbf {\bibinfo {volume}
  {83}},\ \bibinfo {pages} {012304} (\bibinfo {year} {2011})}\BibitemShut
  {NoStop}%
\bibitem [{\citenamefont {Kastoryano}\ \emph {et~al.}(2013)\citenamefont
  {Kastoryano}, \citenamefont {Wolf},\ and\ \citenamefont {Eisert}}]{KWE13}%
  \BibitemOpen
  \bibfield  {author} {\bibinfo {author} {\bibfnamefont {M.~J.}\ \bibnamefont
  {Kastoryano}}, \bibinfo {author} {\bibfnamefont {M.~M.}\ \bibnamefont
  {Wolf}}, \ and\ \bibinfo {author} {\bibfnamefont {J.}~\bibnamefont
  {Eisert}},\ }\href {\doibase 10.1103/PhysRevLett.110.110501} {\bibfield
  {journal} {\bibinfo  {journal} {Phys. Rev. Lett.}\ }\textbf {\bibinfo
  {volume} {110}},\ \bibinfo {pages} {110501} (\bibinfo {year}
  {2013})}\BibitemShut {NoStop}%
\bibitem [{\citenamefont {Michnicki}(2015)}]{Michnicki}%
  \BibitemOpen
  \bibfield  {author} {\bibinfo {author} {\bibfnamefont {K.}~\bibnamefont
  {Michnicki}},\ }\emph {\bibinfo {title} {Towards self-correcting quantum
  memories}},\ \href@noop {} {Ph.D. thesis} (\bibinfo {year}
  {2015})\BibitemShut {NoStop}%
\bibitem [{\citenamefont {{Avis}}(1983)}]{Avi83}%
  \BibitemOpen
  \bibfield  {author} {\bibinfo {author} {\bibfnamefont {D.}~\bibnamefont
  {{Avis}}},\ }\href {\doibase 10.1002/net.3230130404} {\bibfield  {journal}
  {\bibinfo  {journal} {Networks}\ }\textbf {\bibinfo {volume} {13}},\ \bibinfo
  {pages} {475} (\bibinfo {year} {1983})}\BibitemShut {NoStop}%
\bibitem [{\citenamefont {{Yoshida}}(2011)}]{Yos11}%
  \BibitemOpen
  \bibfield  {author} {\bibinfo {author} {\bibfnamefont {B.}~\bibnamefont
  {{Yoshida}}},\ }\href {\doibase 10.1016/j.aop.2011.06.001} {\bibfield
  {journal} {\bibinfo  {journal} {Ann. Phys.}\ }\textbf {\bibinfo {volume}
  {326}},\ \bibinfo {pages} {2566} (\bibinfo {year} {2011})},\ \Eprint
  {http://arxiv.org/abs/1103.1885} {arXiv:1103.1885} \BibitemShut {NoStop}%
\bibitem [{\citenamefont {Pastawski}\ \emph {et~al.}(2010)\citenamefont
  {Pastawski}, \citenamefont {Kay}, \citenamefont {Schuch},\ and\ \citenamefont
  {Cirac}}]{PastawskiLimitations}%
  \BibitemOpen
  \bibfield  {author} {\bibinfo {author} {\bibfnamefont {F.}~\bibnamefont
  {Pastawski}}, \bibinfo {author} {\bibfnamefont {A.}~\bibnamefont {Kay}},
  \bibinfo {author} {\bibfnamefont {N.}~\bibnamefont {Schuch}}, \ and\ \bibinfo
  {author} {\bibfnamefont {I.}~\bibnamefont {Cirac}},\ }\href@noop {}
  {\bibfield  {journal} {\bibinfo  {journal} {Quant. Inf. Comp.}\ }\textbf
  {\bibinfo {volume} {10}},\ \bibinfo {pages} {580} (\bibinfo {year}
  {2010})}\BibitemShut {NoStop}%
\bibitem [{\citenamefont {Haah}(2013)}]{HaahModules}%
  \BibitemOpen
  \bibfield  {author} {\bibinfo {author} {\bibfnamefont {J.}~\bibnamefont
  {Haah}},\ }\href@noop {} {\bibfield  {journal} {\bibinfo  {journal} {Commun.
  Math. Phys.}\ }\textbf {\bibinfo {volume} {324}},\ \bibinfo {pages} {351}
  (\bibinfo {year} {2013})}\BibitemShut {NoStop}%
\bibitem [{\citenamefont {Brell}(2016)}]{Bre14}%
  \BibitemOpen
  \bibfield  {author} {\bibinfo {author} {\bibfnamefont {C.~G.}\ \bibnamefont
  {Brell}},\ }\href {\doibase 10.1088/1367-2630/18/1/013050} {\bibfield
  {journal} {\bibinfo  {journal} {New J. Phys.}\ }\textbf {\bibinfo {volume}
  {18}},\ \bibinfo {pages} {013050} (\bibinfo {year} {2016})}\BibitemShut
  {NoStop}%
\bibitem [{\citenamefont {Lidar}\ and\ \citenamefont
  {Brun}(2013)}]{LidarBrunQuantumErrorCorrection}%
  \BibitemOpen
  \bibfield  {author} {\bibinfo {author} {\bibfnamefont {D.~A.}\ \bibnamefont
  {Lidar}}\ and\ \bibinfo {author} {\bibfnamefont {T.~A.}\ \bibnamefont
  {Brun}},\ }\href@noop {} {\emph {\bibinfo {title} {Quantum error
  correction}}}\ (\bibinfo  {publisher} {Cambridge University Press},\ \bibinfo
  {address} {Cambridge},\ \bibinfo {year} {2013})\BibitemShut {NoStop}%
\end{thebibliography}
%

\end{document}